\documentstyle[12pt,aaspp4,psfig,amstex]{article}
\def\etal{{\it et al.}}    
\def\degr{\hbox{$^\circ$}}

\def\lsim{\mathrel{\hbox{\rlap{\lower.55ex \hbox {$\sim$}}\kern-.0em\raise.4ex \hbox{$<$}}}} 
\def\gsim{\mathrel{\hbox{\rlap{\lower.55ex \hbox {$\sim$}}\kern-.0em\raise.4ex \hbox{$>$}}}}

\def\grb{GRB\thinspace{000418}}
\def\ts{\thinspace}

\begin{document}

\title{GRB\thinspace{000418}: A Hidden Jet Revealed?}

\author{ 
 E. Berger\altaffilmark{1},
 A. Diercks\altaffilmark{1},
 D. A. Frail\altaffilmark{2}, 
 S. R. Kulkarni\altaffilmark{1},
 J. S. Bloom\altaffilmark{1},
 R. Sari\altaffilmark{3},
 J. Halpern\altaffilmark{4},
 N. Mirabal\altaffilmark{4},
 G. B. Taylor\altaffilmark{3},
 K. Hurley\altaffilmark{5},
 G. Pooley\altaffilmark{6},
 K. M. Becker\altaffilmark{7},
 R. M. Wagner\altaffilmark{8},
 D. M. Terndrup\altaffilmark{8},
 T. Statler\altaffilmark{9},
 E. Mazets\altaffilmark{10},
 T. Cline\altaffilmark{11}
}

\altaffiltext{1}{California Institute of Technology, Palomar
 Observatory 105-24, Pasadena, CA 91125}

\altaffiltext{2}{National Radio Astronomy Observatory, P.~O.~Box O,
  Socorro, NM 87801}

\altaffiltext{3}{California Institute of Technology, Theoretical
 Astrophysics 103-33, Pasadena, CA 91125}

\altaffiltext{4}{Astronomy Department, Columbia University 550 West
  120th St., New York, NY 10027}

\altaffiltext{5}{University of California, Berkeley, Space Sciences
  Laboratory, Berkeley, CA 94720-7450}

\altaffiltext{6}{Mullard Radio Astronomy Observatory, Cavendish
  Laboratory, Madingley Road, Cambridge CB3 0HE}      

\altaffiltext{7}{Department of Physics, Oberlin College Oberlin, OH
  44074}

\altaffiltext{8}{Ohio State University, Department of Astronomy, 
Columbus, OH, 43210}

\altaffiltext{9}{Ohio University, Department of Physics and Astronomy, 
Athens, OH, 45701}

\altaffiltext{10}{Ioffe Physico-Technical Institute, St. Petersburg,
  194021 Russia}

\altaffiltext{11}{NASA Goddard Space Flight Center, Code 661,
  Greenbelt, MD 20771}

\begin{abstract}
  We report on optical, near-infrared and centimeter radio
  observations of \grb\ which allow us to follow the evolution of the
  afterglow from 2 to 200 days after the $\gamma$-ray burst. In
  modeling these broad-band data, we find that an isotropic explosion
  in a constant density medium is unable to simultaneously fit both
  the radio and optical data. However, a jet-like outflow with an
  opening angle of 10-20\degr\ provides a good description of the
  data.  The evidence in favor of a jet interpretation is based on the
  behavior of the radio light curves, since the expected jet break is
  masked at optical wavelengths by the light of the host galaxy. We
  also find evidence for extinction, presumably arising from within
  the host galaxy, with A$^{host}_{\rm{V}}$=0.4 mag, and host flux
  densities of $F_R=1.1$ $\mu$Jy and $F_K=1.7$ $\mu$Jy.  These values
  supercede previous work on this burst due to the availability of a
  broad-band data set allowing a global fitting approach. A model in
  which the GRB explodes into a wind-stratified circumburst medium
  cannot be ruled out by these data.  However, in examining a sample
  of other bursts (e.g. GRB\thinspace{990510}, GRB\thinspace{000301C})
  we favor the jet interpretation for \grb.
\end{abstract} 



\section{Introduction}\label{sec:intro}

\grb\ was detected on April 18, 2000, at 09:53:10 UT by the {\em
  Ulysses}, {\em KONUS-Wind} and {\em NEAR} spacecraft, which are part
of the third interplanetary network (IPN) .  The event lasted $\sim$30
s, and a re-analysis of the early Ulysses data (\cite{hcm00}) gives a
fluence of $4.7\times 10^{-6}$ erg cm$^{-2}$ in the 25-100 keV band.
A fit to the total photon spectrum from the KONUS data
$2\times10^{-5}$ erg cm$^{-2}$ in the energy range 15 - 1000 keV.
Intersecting IPN annuli resulted in a 35 arcmin$^2$ error box, in
which Klose \etal\ (2000b\nocite{ksg+00}) identified a variable
near-infrared (NIR) source. The early R-band light curve of this
source was described by Mirabal \etal\ (2000\nocite{mhw+00b}) as
having a power-law decay $t^{-0.84}$, typical for optical afterglows.
The redshift for the host galaxy of $z \simeq 1.119$ was measured by
Bloom \etal\ (2000\nocite{bdd+00}) from an [OII] emission line
doublet.  Assuming cosmological parameters of $\Omega_M$=0.3,
$\Lambda_0$=0.7 and H$_0$=65~km s$^{-1}$~Mpc$^{-1}$, this redshift
corresponds to a luminosity distance d$_{\rm{L}}=2.5\times{10}^{28}$
cm and gives an implied isotropic $\gamma$-ray energy release of
E$_\gamma=1.7\times{10}^{52}$ erg.

Klose \etal\ (2000b\nocite{ksm+00}) have recently summarized
optical/NIR data observations of \grb. In this paper we present
additional optical/NIR data and a complete set of radio observations
between 1.4 GHz and 22 GHz, from 10 to 200 days after the burst. We
use this broad band data set to fit several models, deriving the
physical parameters of the system.

\section{Observations}\label{sec:obs}

\subsection{Optical Observations}\label{sec:optical}
In Table~\ref{tab:optical} we present deep optical photometry obtained
at Palomar, Keck\footnotemark\footnotetext{The W.~M.~Keck Observatory
  is operated by the California Association for Research in Astronomy,
  a scientific partnership among California Institute of Technology,
  the University of California and the National Aeronautics and Space
  Administration.}, and MDM observatories covering six weeks following
the GRB as well as data from the extant literature.

All of the optical data was overscan corrected, flat-fielded, and
combined in the usual manner using IRAF (\cite{iraf}).  PSF-fitting
photometry was performed relative to several local comparison stars
measured by Henden (2000\nocite{h00}) using DoPhot (\cite{dophot}).
Short exposures of the field in each band
were used to transfer the photometry (\cite{h00}) to several fainter
stars in the field.  
Several of the Keck+ESI measurements, and the Palomar 200''
measurement were made in Gunn-r and Gunn-i respectively and were
calibrated by transforming the local comparison stars to the Gunn
system using standard transformations (\cite{wade79},
\cite{jorgensen94}).  We add an additional 5\% uncertainty in
quadrature with the statistical uncertainties to reflected the
inherent imprecision in these transformations.

The Ks-band image of the field was obtained on the Keck I Telescope on
Manua Kea, Hawaii with the Near Infrared Camera.
We obtained a total of 63 one-minute exposures which we reduced and
combined with the {\tt IRAF/DIMSUM} package modified by D.~Kaplan.
There was significant cloud and cirrus cover and so the night was not
photometric.

The HST STIS/Clear image was obtained on 4 June 2000 UT as part of
the TOO program \# 8189 (P.I.~A.~Fruchter) and made public on 2 September
2000 UT.  Five images of 500 s each were obtained which we combined using
the {\tt IRAF/DITHER} task.  The final plate scale is 25 milliarcsec
pixel$^{-1}$.

We corrected all optical measurements in Table~\ref{tab:optical} for a
Galactic foreground reddening of $E(B-V)=0.032$ (\cite{sfd98}) at the
position of the burst $(l,b)=(261.16, 80.78)$ before converting to
flux units (\cite{fsi95}, \cite{bb88}) assuming R$_{\rm V}$=3.1.

\subsection{Radio Observations}\label{sec:vla}
Radio observations were undertaken at a frequency of 15 GHz with the
Ryle Telescope. All other frequencies were observed with either the
NRAO\footnotemark\footnotetext{The NRAO is a facility of the National
  Science Foundation operated under cooperative agreement by
  Associated Universities, Inc.  NRAO operates the VLA and the VLBA}
Very Large Array (VLA) or the Very Long Baseline Array (VLBA).  A log
of these observations can be found in Table \ref{tab:Table-VLA}. The
data acquisition and calibration for the Ryle and the VLA were
straightforward (see \cite{fbg+00} for details).

The single VLBA observation was carried out at 8.35 GHz with a total
bandwidth of 64 MHz in a single polarization using 2 bit sampling for
additional sensitivity.  The nearby ($<$1.3\degr) calibrator
J1224+2122 was observed every 3 minutes for delay, rate and phase
calibration.  Amplitude calibration was obtained by measurements of
the system temperature in the standard way.  The coordinates for \grb\ 
derived from the VLBA detection are (epoch J2000) $\alpha$\ =\ 
$12^h25^m19.2840^s$ ($\pm{0.015^s}$) $\delta$\ =\ 
$+20^\circ06^\prime11.141^{\prime\prime}$
($\pm{0.001^{\prime\prime}}$).

\section{The Optical Light Curve and Host Galaxy}\label{sec:lightc}
In Figure \ref{fig:sixlc} we display the R and K-band light curves
constructed from measurements in Table~\ref{tab:optical}. The
pronounced flattening of the R-band light curve at late times is
reasonably attributed to the optical afterglow fading below the
brightness of the underlying host galaxy. A noise-weighted least
squares fit was made to the data of the form $f_R\ =\ f_o\ts
t^\alpha_o$ + $f_{host}$ for which we derive $f_o=23.4\pm{2.1}$
$\mu$Jy, $\alpha_o=-1.41\pm0.08$ and $f_{host}=1.08\pm{0.06}$ $\mu$Jy
with a reduced $\chi^2_r=0.94$.  Our inferred R-band magnitude for the
host galaxy R$_{host}=23.66\pm0.06$ is nearly identical to that
obtained from a similar analysis by Klose \etal\ 
(2000b\nocite{ksm+00}). In order to estimate the effect of the host in
other optical bands we scaled R$_{host}$ for \grb\ to a spectrum of
the host galaxy of GRB\ts{980703} (\cite{bfk+98}) ($z = 0.966$) whose
magnitude was measured in seven broad-band colors (B, V, R, I, J, H,
and K). Our results indicate that 50-100\% of the flux in some bands
is due to the host galaxy after the first 10 days. Therefore, for the
afterglow modeling in \S{\ref{sec:global}} we chose not to include the
late-time measurements of \grb\ in the B, V, and Gunn-i bands.

\section{The Radio Light Curves}\label{sec:rlightc}

In Figure \ref{fig:sixlc} we display the radio light curves at 4.86,
8.46, 15 and 22 GHz.  To first order all four frequencies show a
maximum near 1 mJy on a time scale of 10 to 20 days. There is no
discernible rising trend at any frequency. This is most clear at 8.46
GHz, where beginning 10 days after the burst, the light curve
undergoes a steady decline, fading from 1 mJy to 0.1 mJy over a six
month period. The temporal slope of the 8.46 GHz light curve after the
first two months $\alpha_{\rm rad}=-1.37\pm0.10$ ($\chi^2_r=1.4$) is similar
to the optical R-band curve $\alpha_{\rm opt}=-1.41\pm0.08$.

Superimposed on this secular decrease, there exist point-to-point
variability of order 50\%, especially in the early measurements. We
attribute these variations to interstellar scintillation (ISS)
(\cite{goo97}, \cite{wal97}). The method by which we estimate the
magnitude of the intensity fluctuations induced by ISS as a function
of frequency and time is described in full by Berger
\etal~(2000\nocite{bsf+00}).  Briefly, we estimate the magnitude of
scattering with the model of Taylor \& Cordes (1993\nocite{tc93}), and
use this to calculate the transition frequency $\nu_0$ between weak
and strong scattering using Walker (1997). The normalizations used in
Goodman (1997) give slightly larger values of $\nu_0$.

In the direction toward \grb\ we derive $\nu_0\simeq{3.6}$ GHz and
therefore most of our measurements were taken in the weak ISS regime.
In this case the modulation scales as $\nu^{-17/12}$, with a maximum of
65\% expected at 4.86 GHz and 30\% at 8.46 GHz. At 15 GHz and 22 GHz
we estimate that the ISS-induced fluctuations are only a fraction of
the instrumental noise. The expansion of the fireball will eventually
quench ISS when the angular size of the fireball exceeds the angular
size of the first Fresnel zone at the distance of the scattering
screen.  The fireball size, and hence the quenching timescale, is
model-dependent, and we use the derived total energy and density from
the global fits (see \S{\ref{sec:global}} below) to estimate this time
for each model.  For example, in a simple spherical fireball this
occurs after 15 days at 4.86 GHz and 10 days at 8.46 GHz, and
thereafter the modulation indices decline as $t^{-35/48}$.  We note
that the observed fluctuations at 4.86 and 8.46 GHz conform to the
predicted level of ISS, but that the measurements at 8.46 GHz from
around 50 days after the burst deviate by a factor of three from the
predicted ISS level.

In addition, we use the scintillation pattern to estimate the true
$\chi^2_r$ for each model, by adding in quadrature to the instrumental
noise an additional ISS-induced uncertainty, $\sigma_{\rm
  ISS}=m_pF_{\nu,{\rm model}}$, where $m_p$ and $F_{\nu,{\rm model}}$
are the modulation index and model flux density at frequency $\nu$,
respectively (Berger \etal\ 2000).

\section{Global Model Fits}\label{sec:global}

The optical and radio data presented here have allowed us to track the
evolution of the \grb\ afterglow from 2 to 200 days after the burst.
With careful modeling of the light curves, it should be possible to
infer the physical parameters of the blast wave and thereby gain some
insight into the nature of GRB progenitors.  In particular, the
hydrodynamic evolution of the shock is governed by the energy of the
explosion, the geometry of the expanding ejecta shock and the type of
environment into which the GRB explodes (\cite{spn98}, \cite{wg99},
\cite{cl99}, \cite{pk00}).  We will consider three basic models: a
spherical explosion in a constant density medium, collimated ejecta
({\it i.e.}, jets) interacting with a constant density medium, and a
spherical explosion in a wind-blown medium.

The starting point for any afterglow interpretation is the
cosmological fireball model (e.g. \cite{mr97a}, \cite{wax97a}). A
point explosion of energy E$_0$ expands relativistically into the
surrounding medium (with density $\rho\propto{r}^{-s}$, where $s=0$
for constant density ISM and $s=2$ for a wind) and the shock produced
as a result of this interaction is a site for particle acceleration.
The distribution of electrons is assumed to be a power-law of index
$p$, and the fraction of the shock energy available for the electrons
and the magnetic field is $\epsilon_e$ and $\epsilon_B$, respectively.
The values of these three quantities ($p$, $\epsilon_e$ and
$\epsilon_B$) are determined by the physics of the shock and the
process of particle acceleration and in the absence of detailed
understanding are taken to be constant with time.

The instantaneous broad-band synchrotron spectrum can be uniquely
specified by the three characteristic frequencies $\nu_{a}$,
$\nu_{m}$, and $\nu_{c}$ ({\it i.e.}, synchrotron self-absorption,
synchrotron peak, and cooling), the peak flux density $f_m$, and $p$.
For this work we adopt the smooth spectral shape as given by Granot
\etal~(1999a\nocite{gps99b}, 1999b\nocite{gps99c}) rather than the
piecewise, broken power-law spectrum used by other authors ({\it
  e.g.}, \cite{wg99}).  The evolution of the spectrum (and thus the
time dependence of $\nu_{a}$, $\nu_{m}$, $\nu_{c}$ and $f_m$) is
governed by the geometry of the explosion (spherical or a collimated
into a jet-like outflow), and the properties of the external
environment (constant density or a radial density profile).  Our
approach is to adopt a model (sphere, wind, jet, etc.)  and solve for
the above spectral parameters using the entire optical and radio data
set.  The advantages and details of global model fitting are discussed
by Berger \etal\ (2000)\nocite{bsf+00}.

\begin{figure*} 
  \centerline{\hbox{\psfig{figure=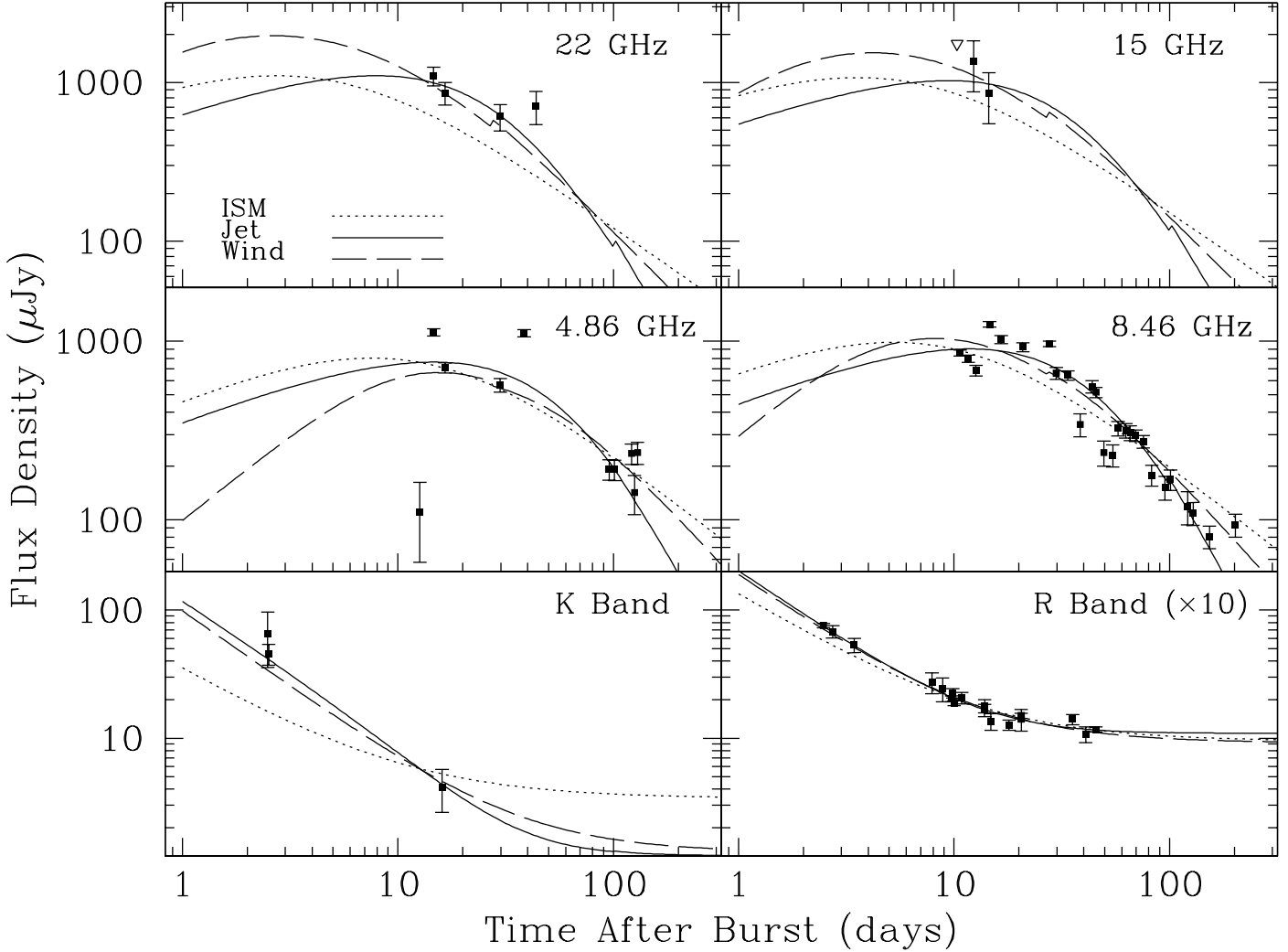,angle=270,width=20cm}}}
\caption[]{Radio and optical light curves for \grb. The observing 
  frequency (or band) is shown in the upper right corner of each
  panel. Optical magnitudes were first corrected for Galactic
  forground reddening before converting to flux units. For display
  purposes the R band flux densities have been increased by a factor
  of 10.  The 8.46 GHz measurements on August 25 and September 18 are
  3-epoch averages taken over a period of 7 days and 15 days,
  respectively.  The dotted, solid and dashed lines are light curves
  for ISM, jet and wind models, respectively. They were derived from a
  global fit to the entire broad-band dataset. See text for more
  details.
\label{fig:sixlc}}
\end{figure*}

The simplest model is a spherically symmetric explosion in a constant
density medium ({\it ISM}: Sari \etal\ 1998). The total $\chi^2_r$ for
this model (see Table~\ref{tab:models}) gives a highly unsatisfactory
fit to the data. On close inspection (Figure \ref{fig:sixlc}) we find
that the model systematically underpredicts the optical flux. Adding
extinction from the host galaxy only makes this worse.  The
fundamental difficulty with the {\it ISM} model is that it predicts
$f_m=constant$, independent of frequency. In this case, since it is
the radio data that is responsible for defining the peak of spectrum,
it results in a value of $f_m$ that is too low at higher frequencies.

To obtain better fits to the joint optical and radio data sets we look
to models for which $f_m$ is time-dependent.  One such model is a
collimated outflow into a medium with uniform density ({\it Jet}:
\cite{rho97b}, \cite{rho99}, \cite{sph99}). The clearest observational
signature of the {\it Jet} model is an achromatic break in the light
curves at $t_{j}$ ({\it e.g.}, \cite{hbf+99}). At radio wavelengths
({\it i.e.}, below $\nu_m$) at $t_{j}$ we expect a transition from a
rising $t^{1/2}$ light curve to a shallow decay of $t^{-1/3}$, while
at optical wavelengths the decay is expected to steepen to $t^{-p}$.
These decay indices refer to the asymptotic values.

Detecting a jet transition at optical wavelengths may be difficult if
it occurs on timescales of a week or more. In these cases the
afterglow is weak and the light from the host galaxy may start to
dominate the light curve ({\it e.g.}, Halpern et
al.~2000\nocite{hum+00}). In such instances radio observations may be
required to clarify matters, since the radio flux is increasing prior
to $t_j$ and changes in the lightcurve evolution due to the jet break
are easily detected.  Indeed, the jet in GRB\ts{970508}, which was
very well observed in the radio is not discernible in the optical
data.  In this case, Frail, Waxman \& Kulkarni (2000)\nocite{fwk00}
found a wide-angle jet with an opening angle of 30\degr\ and $t_j \sim
30$ days (but see \cite{cl00}).

A {\it Jet} model with $t_{j}\approx 26$ days fits the data remarkably
well (see Figure \ref{fig:sixlc}). The strongest point in favor of the
{\it Jet} model is that it reproduces the broad maximum ($\sim$1 mJy)
seen from 5 GHz to 22 GHz. We expect such a plateau at $t_{j}$ as
all light curves for $\nu_a<\nu\leq\nu_m$ reach their peak fluxes
(with only a weak $\nu^{1/3}$ frequency dependence) before undergoing
a slow decline.  Most other models predict a strong frequency
dependence in peak flux which is not seen in this case. 

Knowing $t_{j}$ and the density of the ambient medium n$_0$ from the
model fit (Table \ref{tab:models}) we can make a geometric correction
to the total isotropic energy E$_\gamma$, as determined from either
the observed $\gamma$-ray fluence or the total energy of the afterglow
E$_{52}$, from the fit to the afterglow data (Sari
\etal~1999\nocite{sph99}, \cite{lw00}). This approach gives values for
the jet opening angle $\theta_j$ between 10\degr\ and 20\degr, which
for a two-sided jet reduces the GRB energy to $\sim {10}^{51}$ erg.
The rapid lateral expansion of the jet also accelerates the transition
to the non-relativistic regime, resulting in a change in the evolution
of the light curves. Since this occurs on a timescale
$t_{\rm{NR}}\sim{t_{j}}\theta_{j}^{-2}\sim 350$ days (\cite{wkf98}),
we do not expect the non-relativistic transition to be important for
our data.

There is some freedom in our choice of $\nu_c$. We know that a cooling
break ({\it i.e.}, $\Delta\alpha=-0.25$) is not apparent in the R band
light curve on timescales of 2-10 days, so we searched for solutions
with $\nu_c$ above or below this frequency. We found that physically
consistent solutions ({\it i.e.}, with non-negative host fluxes, and
$\epsilon_B<1$) were only possible for values of $\nu_c$ below
the optical band.

As part of the fitting process we also solved for the
host flux density in the R and K bands and for any local dust
obscuration, assuming an LMC-like extinction law. This yields
$f_{host}$(R)=1.1 $\mu$Jy, $f_{host}$(K)=1.7 $\mu$Jy and
A$^{host}_{\rm{V}}$=0.4 (in the host galaxy restframe). Klose
\etal~(2000b) argued for significant dust extinction with
A$^{host}_{\rm{V}}$=0.96. However, they likely overestimated
A$^{host}_{\rm{V}}$ since they assumed a spherical fireball model and
arbitrarily located $\nu_c$ above the optical band.  Moreover, we find
that there is some covariance between the values of A$^{host}_{\rm{V}}$
and $p$ so that only with a global fit, in which $p$ is constrained by the
radio data as well as the optical data, we can solve for A$^{host}_{\rm{V}}$
in a self-consistent manner.

In view of the claims linking GRBs with the collapse of massive stars
(\cite{gvv+98}, \cite{bkd+99}, \cite{rei99}, \cite{pgg+00}), we
considered a final model of a spherical explosion into a wind-blown
circumburst medium ({\it Wind}: \cite{cl99}, \cite{lc99}). The {\it
  Wind} model (Figure \ref{fig:sixlc}) fits the data as well as the
{\it Jet} model.  In fact, the two models have sufficiently similar
$\chi^{2}$ to be indistinguishable. The close match between the
temporal slopes of the late-time 8.46 GHz light curve and the early R
band light curve (see \S{\ref{sec:rlightc}}) is a point in favor of
the {\it Wind} model since a steeper decline is expected for a jet
geometry. Our failure to distinguish between these two models can be
blamed on the absence of radio measurements (particularly at
millimeter wavelengths) at early times. The rapid rise of the flux
density below $\nu_a$ and $\nu_m$ in the {\it Wind} model and the
strong frequency dependence of the peak flux (see Figure
\ref{fig:sixlc}), make such measurements advantageous (\cite{pk00}).
Moreover, in principle the {\it Wind} model can be distinguished from
the other models by the fact that in this model $\nu_c$ is increasing
with time ($\nu_c \propto t^{1/2}$).  However, in this case since
$\nu_c$ lies below the optical/IR bands, this behavior would be
distinguishable only at late time when the host flux dominates over
the OT.  As before we solved for the host flux and any dust
extinction, yielding $f_{host}$(R)=0.9 $\mu$Jy, $f_{host}$(K)=1.3
$\mu$Jy and A$^{host}_{\rm{V}}$=0.3. In our view further (and more
sophisticated) model fits are not warranted by these data.

In summary, we find that the radio and optical/NIR observations of the
afterglow emission from \grb\ can be fit by two different models. The
close similarity between the results of the {\it Wind} and {\it Jet}
models has been noted for other GRBs: GRB 970508 (Frail \etal\ (2000),
\cite{cl00}), GRB 980519 (\cite{fks+99}, \cite{cl99}), GRB 000301C
(\cite{bsf+00}, \cite{lc00}), and GRB 991208 (\cite{gbb+00},
\cite{lc00}). The resolution of this conflict is important, since it
goes to the core of the GRB progenitor issue. If the GRB progenitor is
a massive star then there must be evidence for a density gradient in
the afterglow light curves, reflecting the stellar mass loss that
occurs throughout the star's lifetime (\cite{cl99}, \cite{pk00}). At
present, an unambiguous case for a GRB afterglow expanding into a wind
has yet to be found. On the contrary, most afterglows are better fit
by a jet expanding in a constant density medium ({\it e.g.},
\cite{hbf+99}, \cite{hum+00}, \cite{pk00b}) and thus we are faced with
a peculiar situation. While there is good evidence linking GRBs to the
dusty, gas-rich environments favored by hypernova progenitors
(\cite{bkd00}, \cite{gw00}), the expected mass loss signature is
absent (or at best ambiguous) in all afterglows studied to date.

\acknowledgements A. Diercks is supported by a Millikan Fellowship at
Caltech.  GRB research at Caltech is supported by NSF and NASA grants
(SRK, SGD, FAH). KH is grateful for Ulysses support under JPL Contract
958056, and for NEAR support under NAG 5 9503.



\begin{deluxetable}{lccccc}
\footnotesize \tablecaption{Optical/NIR Observations of \grb
\label{tab:optical}} \tablewidth{0in} \tablehead{ \colhead{UT Date} &
\colhead{Instr.\tablenotemark{a}} & \colhead{Band} & 
\colhead{Mag.\tablenotemark{b}} &
\colhead{Err.} & \colhead{Ref.\tablenotemark{c}} }
\tablecolumns{7} 
\startdata
Apr 20.89  & TNG 3.5m    & R            & 21.54 & 0.04  & 2   \nl
Apr 20.90  & CA 3.5m    & K$^{\prime}$ & 17.49 & 0.5   & 2   \nl
Apr 20.93  & CA 1.2m    & K$^{\prime}$ & 17.89 & 0.2   & 2   \nl
Apr 21.15  & MDM 2.4m   & R            & 21.66 & 0.12  & 1 \nl 
Apr 21.86  & LO 1.5m    & R            & 21.92 & 0.14  & 2   \nl 
Apr 26.316 & USNO 1.3m  & R            & 22.65 & 0.20  & 2,4 \nl
Apr 27.26  & MDM        & R            & 22.77 & 0.23  & 1 \nl
Apr 28.170 & P200       & R            & 22.97 & 0.06  & 1   \nl 
Apr 28.3   & MDM        & R            & 22.86 & 0.09  & 1   \nl 
Apr 28.413 & Keck/ESI   & R            & 23.05 & 0.05  & 1   \nl
Apr 29.26  & MDM        & R            & 22.95 & 0.11  & 1   \nl
May  2.274 & Keck/ESI   & Gunn-i       & 23.38 & 0.05  & 1   \nl
May  2.28  & MDM        & R            & 23.19 & 0.12  & 1   \nl
May  2.285 & Keck/ESI   & B            & 24.31 & 0.08  & 1   \nl
May  2.31  & USNO 1.3m  & R            & 23.11 & 0.130 & 2   \nl
May  3.26  & USNO 1.3m  & R            & 23.41 & 0.160 & 2   \nl
May  4.44  & UKIRT 3.8m & K            & 20.49 & 0.40  & 2   \nl
May  6.42  & Keck/LRIS  & R            & 23.48 & 0.10  & 7   \nl
May  8.89  & TNG        & R            & 23.30 & 0.05  & 2   \nl
May  8.92  & TNG        & V            & 23.92 & 0.07  & 2   \nl
May  9.82  & USNO 1.0m  & R            & 23.37 & 0.21  & 2   \nl
May 23.93  & TNG        & R            & 23.37 & 0.10  & 2   \nl
May 29.228 & P200       & R            & 23.66 & 0.15  & 1   \nl
Jun 2.88   & CA 3.5m    & R            & 23.32 & 0.08  & 2   \nl
Jun 2.91   & TNG        & R            & 23.57 & 0.05  & 2   \nl
\enddata
\tablenotetext{a}{CA 3.5m=Calar Alto 3.5-meter, USNO1.3m = U.S. Naval
  Observatory Flagstaff Station 1.3-meter, ESI=W.M. Keck Observatory
  Echellette Spectrograph-Imager, LRIS=W.M. Keck Observatory
  Low-Resolution Imaging Spectrograph} \tablenotetext{b}{Optical
  photometry is on the Kron-Cousins and Gunn systems and referred to
  that of Henden (2000\nocite{h00}.) Data are corrected for Galactic
  extinction corresponding to $E(B-V) = 0.032$ derived from the maps
  of Schlegel \etal\ (1998\nocite{sfd98}).}
\tablenotetext{c}{1=this work, 2=\cite{ksm+00}, 3=\cite{mhw+00a},
  4=\cite{hhk00}, 5=\cite{mhw+00b}, 6=\cite{mhw+00c}, 7=\cite{mf00}}
\end{deluxetable}


\newpage 
\begin{deluxetable}{lccc}
\tabcolsep0in\footnotesize
\tablewidth{0in}
\tablecaption{Radio Observations of \grb\label{tab:Table-VLA}}
\tablehead {
\colhead {Epoch}      &
\colhead {Telescope} &
\colhead {$\nu_0$} &
\colhead {S$\pm\sigma$} \\
\colhead {(UT)}      &
\colhead {} &
\colhead {(GHz)} &
\colhead {($\mu$Jy)}
}
\startdata
2000 Apr    28.75  & Ryle & 15.0  &  550$\pm$600 \nl
2000 Apr    29.07  &  VLA &  8.46 &  856$\pm$33  \nl
2000 Apr    30.07  &  VLA &  8.46 &  795$\pm$37  \nl 
2000 Apr    30.73  & Ryle & 15.0  & 1350$\pm$480 \nl 
2000 May     1.06  &  VLA &  4.86 &  110$\pm$52  \nl
2000 May     1.06  &  VLA &  8.46 &  684$\pm$48  \nl 
2000 May     2.93  & Ryle & 15.0  &  850$\pm$300 \nl
2000 May     3.04  &  VLA &  4.86 & 1120$\pm$52  \nl
2000 May     3.04  &  VLA &  8.46 & 1240$\pm$46  \nl
2000 May     3.04  &  VLA & 22.46 & 1100$\pm$150 \nl
2000 May     4.97  &  VLA &  1.43 &  210$\pm$180 \nl
2000 May     4.97  &  VLA &  4.86 &  710$\pm$47  \nl
2000 May     4.97  &  VLA &  8.46 & 1020$\pm$53  \nl
2000 May     4.97  &  VLA & 22.46 &  860$\pm$141 \nl
2000 May     7.18  & VLBA &  8.35 &  625$\pm$60  \nl
2000 May     9.25  &  VLA &  8.46 &  926$\pm$53  \nl
2000 May    16.13  &  VLA &  8.46 &  963$\pm$34  \nl
2000 May    18.24  &  VLA &  4.86 &  567$\pm$50  \nl 
2000 May    18.24  &  VLA &  8.46 &  660$\pm$50  \nl         
2000 May    18.24  &  VLA & 22.46 &  610$\pm$114 \nl         
2000 May    22.21  &  VLA &  8.46 &  643$\pm$38  \nl
2000 May    26.92  &  VLA &  4.86 & 1105$\pm$51  \nl         
2000 May    26.92  &  VLA &  8.46 &  341$\pm$50  \nl  
2000 Jun     1.14  &  VLA &  8.46 &  556$\pm$43  \nl         
2000 Jun     1.14  &  VLA & 22.46 &  710$\pm$16  \nl 
2000 Jun     3.04  &  VLA &  8.46 &  517$\pm$34  \nl
2000 Jun     7.01  &  VLA &  8.46 &  238$\pm$38  \nl
2000 Jun    11.93  &  VLA &  8.46 &  230$\pm$33  \nl
2000 Jun    15.13  &  VLA &  8.46 &  325$\pm$30  \nl 
2000 Jun    20.10  &  VLA &  8.46 &  316$\pm$30  \nl 
2000 Jun    23.19  &  VLA &  8.46 &  306$\pm$29  \nl 
2000 Jun    27.08  &  VLA &  8.46 &  296$\pm$22  \nl 
2000 Jul    02.98  &  VLA &  8.46 &  274$\pm$22  \nl 
2000 Jul    10.04  &  VLA &  8.46 &  178$\pm$24  \nl 
2000 Jul    22.81  &  VLA &  8.46 &  152$\pm$23  \nl
2000 Jul    22.81  &  VLA &  4.86 &  192$\pm$25  \nl
2000 Jul    28.50  &  VLA &  8.46 &  168$\pm$22  \nl
2000 Jul    28.50  &  VLA &  4.86 &  191$\pm$25  \nl
2000 Aug    17.74  &  VLA &  8.46 &  119$\pm$25  \nl
2000 Aug    17.74  &  VLA &  4.86 &  235$\pm$31  \nl
2000 Aug    21.65  &  VLA &  4.86 &  142$\pm$35  \nl
2000 Aug    21.65  &  VLA &  8.46 &   87$\pm$31  \nl
2000 Aug    25.78  &  VLA &  4.86 &  238$\pm$34  \nl
2000 Aug    25.78  &  VLA &  8.46 &  166$\pm$27  \nl
2000 Aug    27.89  &  VLA &  8.46 &  100$\pm$25  \nl
2000 Sep    10.73  &  VLA &  8.46 &  148$\pm$25  \nl
2000 Sep    18.68  &  VLA &  8.46 &   55$\pm$20  \nl
2000 Sep    26.62  &  VLA &  8.46 &   85$\pm$22  \nl
2000 Nov    06.55  &  VLA &  8.46 &   94$\pm$14  \nl
\enddata
\tablecomments{The columns are (left to right), (1) UT date of the
  start of each observation, (2) telescope name, (3) observing
  frequency, and (4) peak flux density at the best fit position of the
  radio transient, with the error given as the root mean square noise
  on the image.}
\end{deluxetable}

\begin{deluxetable}{lcccc}
\footnotesize \tablecaption{Model parameters for \grb
\label{tab:models}} \tablewidth{0in} \tablehead{ 
\colhead{Parameters \tablenotemark{a}}  &
\colhead{ISM}        & 
\colhead{Jet}  &
\colhead{Wind}       
}
\tablecolumns{5} 
\startdata
$\nu_{a}$ (Hz)       & 4.1$\times{10}^9$       & 1.7$\times{10}^9$         & 30$\times{10}^9$             \nl
$\nu_{m}$ (Hz)      & 2.3$\times{10}^{11}$    & 1.8$\times{10}^{10}$      & 5.8$\times{10}^{11}$          \nl
$\nu_{c}$ (Hz)       & $2    \times 10^{15}$ & $10^{14}$ & 1.8$\times{10}^{13}$ \nl 
$f_m$ (mJy)           & 2.5          & 3.4         & 10.4           \nl 
$p$             & 2.3          & 2.4         & 2.2            \nl
$t_{jet}$ (days)      & ---          & 25.7          &  ---           \nl
A$^{host}_{\rm{V}}$    & 0.0          & 0.4         & 0.3            \nl \hline
$\chi^{2}$/dof      & 326/54       & 165/54      & 184/54         \nl
\omit           & \omit        & \omit       & \omit          \nl
E$_{52}$        & 11           & 10          & 4              \nl
n$_0$ or A$^*$ & 0.01         & 0.02        & 0.14           \nl
$\epsilon_{B}$  & 0.05         & 0.06        & 0.04           \nl
$\epsilon_{e}$  & 0.03         & 0.10        & 0.07           \nl
\enddata
\tablenotetext{a}{For the {\it ISM} and {\it Wind} models $\nu_{a}$,
  $\nu_{m}$, $\nu_{c}$ and $ f_m$ are the self-absorption, synchrotron
  peak, and cooling frequencies, and the peak flux density,
  respectively on day 1.  For the {\it Jet} model these values are
  referenced instead to the jet break time $t_j$=25.7 d.  $p$ is the
  electron power-law index and A$_{\rm{V}}$ is the V band extinction
  in the rest frame of the host galaxy ($z$=1.118), assuming an
  LMC-like extinction curve.  Note that in the {\it Jet} and {\it ISM}
  models $\nu_c$ was fixed at $10^{14}$ and $2\times 10^{15}$ Hz,
  respectively, but in the {\it Wind} model it was left as a free
  parameter. The resulting values of $\chi^{2}$ include an estimated
  contribution of interstellar scattering (ISS) and the increased
  error in subtracting off a host galaxy flux from each of the optical
  points. The model parameters are the total isotropic energy E$_{52}$
  in units of 10$^{52}$ erg, the ambient density n$_0$ in cm$^{-3}$ or
  in the case of the {\it Wind} model the parameter A$^*$ as defined
  by Chevalier \& Li (1999)\nocite{cl99}.  $\epsilon_{e}$ and
  $\epsilon_{B}$ are the fraction of the shock energy in the electrons
  and the magnetic field, respectively. The true uncertainties in the
  derived parameters are difficult to quantify due to covariance, but
  we estimate that they range from $10-20$\%}
\end{deluxetable}



\end{document}